\documentclass[final,5p,times,twocolumn,numbers,sort&compress]{elsarticle} 
\usepackage{amssymb}
\usepackage{lipsum}
\usepackage{xcolor}
\usepackage{hyperref}   
\usepackage{ulem}
\usepackage{amssymb}
\usepackage{amsmath}
\usepackage{dcolumn}
\usepackage{graphicx}
\usepackage{mathrsfs}
\usepackage{subfigure}
\usepackage{booktabs}
\usepackage{color}
\usepackage{docmute}
\usepackage{hyphenat}
\usepackage{CJKutf8}
\usepackage{multirow}
\usepackage{array}
\usepackage{cancel}
\usepackage{threeparttable}
\usepackage[mathscr]{euscript}

\usepackage{graphicx}
\usepackage{subfigure}
\usepackage{tikz}
\usetikzlibrary{positioning}
\usepackage{xcolor}

\newcommand{\bea}{\begin{eqnarray}}
\newcommand{\eea}{\end{eqnarray}}
\newcommand{\beq}{\begin{equation}}
\newcommand{\eeq}{\end{equation}}

\def\/{\over}

\hypersetup{
    colorlinks = true,
    citecolor  = green,   
    urlcolor   = blue,
    linkcolor  = black
}

\journal{Physics Letters B}

\begin{document}

\begin{frontmatter}

\title{Quantum gravitodiamagnetic interaction}

\author[first]{Di Hao}

\author[first]{Jiawei Hu\corref{cor1}}
\ead{jwhu@hunnu.edu.cn}

\author[first]{Hongwei Yu\corref{cor2}}
\ead{hwyu@hunnu.edu.cn}

\cortext[cor1]{Corresponding author.}
\cortext[cor2]{Corresponding author.}

\affiliation[first]{organization={Department of Physics, Key Laboratory of Low Dimensional Quantum Structures and Quantum Control of Ministry of Education, and Hunan Research Center of the Basic Discipline for Quantum Effects and Quantum Technologies, Hunan Normal University},
            city={Changsha},
            postcode={410081}, 
            state={Hunan},
            country={China}}

\begin{abstract}
In the framework of linearized quantum gravity, we investigate the quantum gravitational interaction induced by the gravitodiamagnetic coupling of two massive objects to vacuum fluctuations of the gravitational field. Starting from the Lagrangian of a particle in a gravitational field and employing the formalism of Weyl gravitoelectromagnetism, we derive the  interaction Hamiltonian  associated with gravitodiamagnetic coupling.  Unlike the linear couplings that arise in gravitoelectric and gravitomagnetic interactions, the gravitodiamagnetic coupling depends quadratically on the gravitomagnetic field. Based on this  Hamiltonian, we show that, for a spherically symmetric gravitational hydrogen-like system in its ground state, the induced quadrupole moment has the opposite sign to the applied gravitomagnetic field, which is the defining signature of gravitodiamagnetism. Using leading-order perturbation theory, we derive an explicit expression for the gravitodiamagnetic interaction potential, valid for $r\gg a$, where $r$ is the separation and $a$ the characteristic size of the objects. Unlike previously studied quantum gravitational interactions, which exhibit distinct power laws in the near and far regions defined relative to the characteristic transition wavelength of the gravitational hydrogen-like systems, the gravitodiamagnetic interaction is attractive and scales as $r^{-11}$ throughout its regime of validity.
\end{abstract}


\end{frontmatter}

\section{Introduction}
\label{introduction}

The interaction between massive bodies via gravity has been a central topic in physics since 1687, when Isaac Newton formulated the law of universal gravitation in his seminal work, \textit{Philosophiae Naturalis Principia Mathematica}~\cite{Newton}.
In Newton's classical theory of gravity, the gravitational potential energy between two massive objects scales as $r^{-1}$, where $r$ is the distance between the two objects. However, this characteristic power-law dependence is expected to be modified when quantum effects are taken into account. 

Although a complete and consistent theory of quantum gravity, capable of describing gravitational phenomena at the Planck scale, remains elusive, quantum corrections to gravity at low energies can be   explored by treating general relativity as an effective field theory. Within this framework, it has been demonstrated  that quantum corrections to the Newtonian potential arise from one-loop diagrams involving off-shell gravitons, leading to additional attractive terms that scale as $r^{-3}$~\cite{Donoghue1994prl,Donoghue1994prd,Hamber1995,Kirilin2002,Holstein2003,Holstein2005}.

Another  fruitful  approach to probing low-energy quantum gravity effects is through linearized quantum gravity. If gravity is inherently quantum,  then one expects fluctuating gravitational fields even in vacuum. 
 These fluctuations can be treated as perturbations propagating over a flat spacetime background, and their behavior can be analyzed using linearized gravity.  In this weak-field regime,  Einstein's field equations can be recast into a form formally analogous to Maxwell's equations, an approach known as Weyl gravitoelectromagnetism~\cite{Campbell1976,Matte1953,Campbell1971,Szekeres1971,Maartens1998,Ruggiero2002,Ramos2010}.  In this analogy,  gravitoelectric and gravitomagnetic fields play roles similar to the electric and magnetic fields, respectively. 
 In such a framework, fluctuating gravitoelectric and gravitomagnetic fields can induce instantaneous mass and mass-current quadrupole moments in massive objects. These induced multipole moments lead to quantum corrections to the Newtonian potential beyond the standard monopole-monopole interaction. 
  In particular, the interaction between two induced mass quadrupole moments, arising from gravitoelectric field fluctuations, has been shown to produce an attractive quantum gravitational potential that scales as $r^{-10}$ in the near-field regime and as $r^{-11}$ in the far-field regime~\cite{Ford2016,Wu2016,Holstein2017}.  A similar scaling behavior has been found for the interaction between induced mass-current quadrupole moments due to gravitomagnetic field fluctuations~\cite{hao2024}. More recently, a mixed interaction between a gravitoelectrically polarizable object and a gravitomagnetically polarizable one has been investigated~\cite{hao2025}. This gravitoelectric-gravitomagnetic interaction, arising from the coupling between an induced mass quadrupole and an induced mass-current quadrupole, results in a positive potential corresponding to a repulsive force that counteracts the classical gravitational attraction. The interaction exhibits a distance dependence that transitions from $r^{-8}$ in the near-field to $r^{-11}$ in the far-field regime.

Yet, the full structure of quantum gravitational multipole interactions is still not completely understood. In electromagnetic theory, the leading terms in the multipolar interaction Hamiltonian include electric dipole, magnetic dipole (paramagnetic), and diamagnetic couplings~\cite{Salam_2000,Buhmann2013,Salam2010}. Unlike the linear electric and magnetic dipole interactions, the diamagnetic term exhibits a quadratic dependence on the magnetic field~\cite{Salam_2000,Buhmann2013,Salam2010}. By analogy, in the gravitoelectromagnetic formalism, the couplings of the mass quadrupole to the gravitoelectric field and the mass-current quadrupole to the gravitomagnetic field resemble electric and magnetic dipole couplings, respectively, and are both linear in the field strength. 
This analogy strongly suggests the existence of a gravitational analog of the diamagnetic interaction, which we refer to as the gravitodiamagnetic coupling.

In this work, we investigate the quantum gravitational interaction between two objects that couple quadratically to the fluctuating gravitomagnetic field, referred to as the quantum gravitodiamagnetic interaction. This interaction represents a new type of quantum correction to the classical Newtonian potential. Specifically, we aim to determine the distance dependence of the interaction energy, and to assess whether the resulting force is attractive or repulsive.

 Throughout the paper,  Greek indices ($\mu, \nu, \ldots$) run from 0 to 3 and denote spacetime components, while Latin indices ($i, j, \ldots$) run from 1 to 3 and denote spatial components. We adopt the Minkowski metric $\eta_{\mu\nu} = \text{diag}(-1, +1, +1, +1)$ and use the Einstein summation convention for repeated indices. Unless otherwise specified, natural units are used with $c = \hbar = 1$.

\section{The interaction Hamiltonian}
\label{sec_ge}

A central theme in modern physics is the interaction between matter and fundamental fields. Electromagnetism provides a paradigmatic example. Its success in describing light-matter interactions, ranging from atomic spectra to Casimir forces, provides a natural reference for exploring analogous phenomena in other contexts.
In the framework of electromagnetism, a neutral atom or molecule $A$ can be modeled as a collection of particles $\mathbb{P} \in A$ with charges $q_{\mathbb{P}}$ (satisfying $\sum_{\mathbb{P} \in A}{q_{\mathbb{P}}=0}$) and masses $m_{\mathbb{P}}$. The leading terms of the multipolar atom–field coupling Hamiltonian are given by~\cite{Salam_2000,Buhmann2013,Salam2010}
\beq \label{ele_mul_Ha}
H^{\mathrm{EM}}_{\mathrm{coup}}=-\mathbf{d}_A\cdot \mathbf{E} -\mathbf{m}_A\cdot \mathbf{B} +\sum_{\mathbb{P} \in A}{\frac{q_{\mathbb{P}}^{2}}{2m_{\mathbb{P}}}}\mathbf{A}^2,
\eeq
where $\mathbf{d}_A$ and $\mathbf{m}_A$ denote the electric and magnetic dipole operators, respectively, and $\mathbf{E}$, $\mathbf{B}$ and $\mathbf{A}$ are the electric field, magnetic field, and electromagnetic vector potential operators. The three terms in Eq.~\eqref{ele_mul_Ha} correspond to electric-dipole, magnetic-dipole, and diamagnetic couplings, respectively. Notably, in contrast to the linear electric- and magnetic-dipole couplings, the diamagnetic coupling
\beq \label{EM_DM}
H_{\mathrm{dia}}=\sum_{\mathbb{P}}{\frac{q^2_{\mathbb{P}}}{2m_{\mathbb{P}}} \mathbf{A} ^2}
\eeq
is inherently quadratic in the vector potential $\mathbf{A}$.

Guided by this well-established framework, we now turn to gravity. In the weak-field limit of general relativity, the spacetime metric $g_{\mu\nu}$ can be expanded around the flat background metric $\eta_{\mu\nu}$ as $g_{\mu\nu}=\eta_{\mu\nu}+h_{\mu\nu}$,
where the perturbation $|h_{\mu\nu}|\ll 1$. In this regime, expanding the Einstein-Hilbert action to linear order in $h_{\mu\nu}$ yields the gravity-matter interaction Lagrangian density $\mathcal{L} = \frac{1}{2} h_{\mu\nu} T^{\mu\nu}$~\cite{MTW}, where $T^{\mu\nu}$ is the energy-momentum tensor of matter. In a local inertial frame described by Fermi normal coordinates, the metric perturbation components $h_{00}$ and $h_{0i}$ can be expanded as Taylor series in powers of the Fermi coordinates, with the expansion coefficients determined by the Riemann curvature tensor. On the other hand, according to the Weyl gravitoelectromagnetism, one defines the gravitoelectric and gravitomagnetic fields as $E_{ij}=-C_{0i0j}$ and $B_{ij}=\frac{1}{2}\epsilon_{ifl}C_{fl0j}$, respectively.
Here, $C_{\alpha\beta\gamma\sigma}$ is the Weyl tensor and $\epsilon_{ifl}$ is the third-order Levi-Civita tensor. In vacuum, the Weyl tensor coincides with the Riemann tensor, allowing the metric perturbations to be expressed in terms of the gravitoelectric field and gravitomagnetic field. This leads to the linear multipole interaction Hamiltonian between the gravitational field and the object~\cite{hao2024,hao2025}
\beq \label{GEandGM}
H_{\mathrm{int}}=-\frac{1}{2}Q^{ij}E_{ij}-\frac{2}{3}S^{ij}B_{ij},
\eeq
where $Q^{ij}$ and $S^{ij}$ are the mass quadrupole and mass-current quadrupole moments, respectively. These terms describe gravitoelectric and gravitomagnetic quadrupole couplings, which are the gravitational analogs of the electric and magnetic dipole interactions in electromagnetism. Here, we note a correction to our previous work: the coefficient of the gravitomagnetic coupling term in Eq. (11) of Ref.~\cite{hao2024} should be  $-\frac{2}{3}$  rather than $-\frac{1}{3}$. This discrepancy arose because only the contribution from the $h_{0i}T^{0i}$ term was included, whereas the symmetric $h_{i0}T^{i0}$ term was omitted.

The formal analogy  outlined  above suggests that the coupling of matter to weak gravitational fields may be organized in a manner similar to the multipolar scheme of electrodynamics. This naturally raises the question of whether a gravitational counterpart to the diamagnetic coupling exists. We refer to such a contribution as the gravitodiamagnetic coupling. 
Here, the metric perturbation component $h_{0i}$ plays the role of a gravitational vector potential, analogous to the electromagnetic vector potential $\mathbf{A}$, as is well established in gravitoelectromagnetism~\cite{Mashhoon_review}. 
This analogy motivates us to consider a contribution proportional to $(h_{0i})^2$.

Typically, the interaction between matter and gravity in the weak-field regime is derived from a linear expansion of the Einstein-Hilbert action in the metric perturbation $h_{\mu\nu}$. While this approach captures the linear gravitoelectric and gravitomagnetic quadrupole couplings, it omits higher-order terms.   
Hence, we start from the fundamental Lagrangian of a massive particle in a gravitational field and derive the corresponding interaction Hamiltonian, taking into account both the quadratic terms in the metric perturbation and the higher-order Fermi-coordinate expansion.
Consider a particle moving in a gravitational field with a speed much less than the speed of light. The Lagrangian of the particle takes the standard form~\cite{MTW,Landau,DeWitt}
\beq \label{Lagrangian}
L=-m\sqrt{-g_{\mu \nu}\dot{x}^{\mu}\dot{x}^{\nu}},
\eeq
where $m$ is the rest mass of the particle, and $x^{\mu}$ is its 4-position. Here, we choose $x^{0}$ as the worldline parameter~\cite{DeWitt}, so that the overdot represents differentiation with respect to the coordinate time $t$. 
The canonical momentum of the particle reads
\beq \label{can_mom}
P_i=\frac{\partial L}{\partial v^i}=\varGamma  m\left( g_{0i}+g_{ij}v^j \right),
\eeq
where $\varGamma =1/\sqrt{-g_{00}-2g_{0i}v^i-g_{ij}v^iv^j}$.
The generalized velocity of the system can be derived from Eq. (\ref{can_mom}), yielding $v^k=\left( \tilde{g}^{ik}P_i \right) /\left( \varGamma m \right)-\tilde{g}^{ik}g_{0i}  $,
where $\tilde{g}^{ik}$ is the inverse spatial metric, defined by $\tilde{g}^{ik}g_{jk}=\delta _{\,\, j}^{i}$, and explicitly given by $\tilde{g}^{ij}=g^{ij}-\left( g^{0i}g^{0j} \right) / {g^{00}}$.
Performing a Legendre transform, one finds that the Hamiltonian can be expressed as
\beq \label{Le_Ham_ex}
H=\sqrt{\tilde{g}^{kl}g_{0k}g_{0l}-g_{00}}\sqrt{m^2+\tilde{g}^{ik}P_iP_k}-\tilde{g}^{ik}g_{0i}P_k.
\eeq

Since we are interested in effects up to second order in the metric perturbation $h_{\mu\nu}$, we expand the inverse metric perturbatively using the defining relation $g^{\mu\sigma}g_{\sigma\nu}=\delta _{\,\, \nu}^{\mu}$. Retaining terms through second order in  $h_{\mu\nu}$ ensures that this relation is satisfied consistently to the same order. The resulting expansion of the inverse metric reads $g^{\mu \nu}=\eta ^{\mu \nu}-h^{\mu \nu}+\eta ^{\alpha \nu}h^{\mu \lambda}h_{\lambda \alpha}$~\cite{Donoghue1994prd,Hooft1974,Holstein2006Am},
and the inverse spatial  metric correspondingly  takes the form $\tilde{g}^{ik}=\eta ^{ik}-h^{ik}+\eta ^{kl}h^{i0}h_{0l}+\eta ^{kl}h^{im}h_{lm}+h^{0i}h^{0k}$.
After subtracting the rest energy, expanding the Hamiltonian (Eq.~(\ref{Le_Ham_ex}))  to second order in \(h_{\mu\nu}\), and retaining the momentum-independent terms in the nonrelativistic limit, we obtain
\begin{equation}\label{eq7}
H_{\mathrm{int}}=
-\frac{1}{2}mh_{00}
+\frac{1}{2}mh_{0i}h_{0i}
-\frac{1}{8}mh_{00}^{2}.    
\end{equation}

First, consider the contribution from $\frac{1}{2}mh_{0i}h_{0i}$. For a gravitationally bound system consisting of several mass points $m_{\mathbb{P}}$, this contribution is
\beq \label{bou_sys_H_GDM}
H^{(0i)}_{\mathrm{GDM}}=\sum_{\mathbb{P}}{\frac{1}{2}m_{\mathbb{P}}\left( h _{0i} \right) ^2}\;.
\eeq
This contribution is the gravitational analog of the electromagnetic diamagnetic Hamiltonian (\ref{EM_DM}) obtained 
by formally replacing the electric charge $q$ with the mass $m$ and the electromagnetic vector potential $\mathbf{A}$ with the gravitational vector potential $h_{0i}$.

In the following, we express the contribution in Eq.~(\ref{bou_sys_H_GDM}) in terms of the gravitomagnetic tensor. 
Since we are interested in the interaction between two objects induced by fluctuating gravitational fields in the steady state over long timescales, it is necessary to adopt a locally inertial coordinate system for an extended time. A suitable choice is the Fermi normal coordinate system~\cite{Boughn2006}. Expanding the metric perturbation component $h_{0i}$ in Eq.~(\ref{bou_sys_H_GDM}) as Taylor series in powers of the Fermi coordinates, we have, to the second order~\cite{MTW,Mashhoon2001,Mashhoon_review,Bini2022,Mashhoon2021(2)},
\beq \label{h0i_ex}  
h_{0i}=-\frac{2}{3}\mathcal{R}_{0jik}x^jx^k.
\eeq
Here, $x^i$ is the spatial coordinate in the Fermi normal coordinate system, and $\mathcal{R}_{\mu\nu\sigma\tau}$ is the Riemann tensor. Comparing Eq.~(\ref{h0i_ex}) with $B_{ij}=\frac{1}{2}\epsilon_{ifl}C_{fl0j}$, and noting that the Riemann tensor coincides with the Weyl tensor in the vacuum case, we can express $h_{0i}$ as 
\beq \label{h0i_new}
h_{0i}=\frac{2}{3}\epsilon _{kil}B_{lj}x_kx_j.
\eeq
Substituting Eq.~(\ref{h0i_new}) into Eq. (\ref{bou_sys_H_GDM}), we obtain
\beq \label{H_0i_GDM}
H^{(0i)}_{\mathrm{GDM}}=\frac{2}{9}\sum_{\mathbb{P}}{m_{\mathbb{P}}\epsilon_{kil}x_{k}^{\mathbb{P}}x_{j}^{\mathbb{P}}\epsilon _{min}x_{m}^{\mathbb{P}}x_{f}^{\mathbb{P}}B_{lj}B_{nf}}.
\eeq

Second, consider the term $-\frac{1}{2}mh_{00}$ in Eq.~\eqref{eq7}. The fourth-order Fermi coordinate expansion of $h_{00}$ contains the curvature-product term $\frac{1}{3}\mathcal{R}^{\mu}{}_{ij0}\mathcal{R}_{\mu kl0}x_i x_j x_k x_l$~\cite{Ishii2005}. Its part quadratic in the gravitomagnetic curvature components $\mathcal{R}_{0jik}$ can be expressed as $\left.h_{00}^{(4)}\right|_{B^2}
=\frac{1}{3}  
\epsilon_{kil}x_kx_j
\epsilon_{min}x_mx_f
B_{lj}B_{nf}$.
Consequently, the term $-mh_{00}/2$ yields
\beq  \label{H_00_GDM}
H_{\mathrm{GDM}}^{(00)}
=-\frac{1}{6}
\sum_{\mathbb{P}}m_{\mathbb{P}}
\epsilon_{kil}x_{k}^{\mathbb{P}}x_{j}^{\mathbb{P}}
\epsilon_{min}x_{m}^{\mathbb{P}}x_{f}^{\mathbb{P}}
B_{lj}B_{nf}.
\eeq

Third, consider the remaining term in Eq.~\eqref{eq7}, $-mh_{00}^{2}/8$. 
Since the leading Fermi-coordinate term in $h_{00}$ is $-\mathcal R_{0i0j}x_ix_j$, this term contributes only to the quadratic gravitoelectric coupling at fourth order in the spatial Fermi coordinates and does not affect the quadratic gravitomagnetic coupling.

Thus, combining the contributions in Eqs. (\ref{H_0i_GDM}) and (\ref{H_00_GDM}), we obtain the gravitodiamagnetic coupling Hamiltonian
\beq   \label{sum_GDM}
H_{\mathrm{GDM}}
=H_{\mathrm{GDM}}^{(0i)}
+H_{\mathrm{GDM}}^{(00)}
=\frac{1}{18}
\sum_{\mathbb{P}}m_{\mathbb{P}}
\epsilon_{kil}x_{k}^{\mathbb{P}}x_{j}^{\mathbb{P}}
\epsilon_{min}x_{m}^{\mathbb{P}}x_{f}^{\mathbb{P}}
B_{lj}B_{nf}.
\eeq

Notably, in contrast to the linear dependence characterizing gravitoelectric and gravitomagnetic couplings, the gravitodiamagnetic coupling exhibits a quadratic dependence on the gravitomagnetic field. Defining  
$\hat{\sigma}_{ljnf}\equiv
\frac{1}{9}
\sum_{\mathbb{P}}
{m_{\mathbb{P}}
\epsilon_{kil}x_{k}^{\mathbb{P}}x_{j}^{\mathbb{P}}
\epsilon_{min}x_{m}^{\mathbb{P}}x_{f}^{\mathbb{P}}}$,
the Hamiltonian \eqref{sum_GDM} can be rewritten as 
\beq \label{gravitodiam_Ham}
H_{\mathrm{GDM}}=\frac{1}{2}\hat{\sigma}_{ljnf}B_{lj}B_{nf}.
\eeq

We now examine the properties of the gravitodiamagnetizability based on the interaction Hamiltonian shown in Eq.~(\ref{gravitodiam_Ham}). 
In the context of quantum mechanics, the gravitationally bound system introduced above can, for simplicity, be modeled as a gravitational hydrogen atom, analogous to the Bohr model of the hydrogen atom. The Hamiltonian of the total system can be written as $H=H_0+H'$,
where $H_0=\sum_{\mathbb{P}}{p_{\mathbb{P}}^{2}/2m_{\mathbb{P}}}+\sum_{\mathbb{P} \ne \mathbb{P}’}{V\left( r_{\mathbb{P}}-r_{\mathbb{P}'} \right)}$ is the unperturbed Hamiltonian of the system with potential $V$, and $H’=H_{\mathrm{GDM}}=\frac{1}{2}\hat{\sigma}_{ljnf}B_{lj}B_{nf}$ is the perturbation due to the gravitodiamagnetic coupling with the applied gravitomagnetic field $B_{ij}$. Then, the corresponding first-order energy shift is
\beq \label{energyshift}
\Delta E_{\mathrm{GDM}}=\langle g|H'|g \rangle ,
\eeq
where $|g\rangle$ denotes the ground state of the unperturbed system. Substituting Eq.~(\ref{gravitodiam_Ham}) into Eq.~(\ref{energyshift}), the energy shift is given by $\Delta E_{\mathrm{GDM}}=\frac{1}{2}\langle g|\hat{\sigma}_{ljnf}|g \rangle B_{lj}B_{nf}$.
From this result, the induced gravitomagnetic quadrupole moment (mass-current quadrupole moment) can be obtained from the definition of the gravitomagnetic coupling shown in Eq. (\ref{GEandGM}) as
\beq \label{GDM_Snf}
S_{nf}= -\frac{3}{2}\frac{\partial \Delta E_{\mathrm{GDM}}}{\partial B_{nf}}=-\frac{3}{2}\langle g|\hat{\sigma}_{ljnf}|g \rangle B_{lj}.
\eeq 
Therefore, $-\langle g|\hat{\sigma}_{ljnf}|g \rangle$ serves as the gravitodiamagnetizability of the ground-state gravitational atom. 
 It is noteworthy that, for a ground-state gravitational hydrogen atom with spherical symmetry, the gravitomagnetic field always induces a quadrupole moment with the opposite sign (see \ref{appd1} for details), reflecting the microscopic mechanism of the gravitodiamagnetic effect.

\section{The quantum gravitodiamagnetic interaction}\label{sec3}

We consider a pair of massive objects A and B coupled with fluctuating gravitational fields in a vacuum. Each object is modeled as a two-level system, with ground and excited states denoted by $|g_{A(B)}\rangle$ and $|e_{A(B)}\rangle$, respectively, and the corresponding energy level spacings are $\omega_{A}$ and $\omega_{B}$, respectively. The total Hamiltonian of the object-field system reads
\beq \label{tot_Ham}
H_{\mathrm{tot}}=H_{A}+H_{B}+H_{F}+H_{AF}+H_{BF},
\eeq
where $H_{F}$ is the Hamiltonian of the fluctuating gravitational field, $H_{A(B)}$ is the Hamiltonian of object A(B), and $H _{A(B)F}$ represents the gravitodiamagnetic interaction Hamiltonian between object A(B) and the fluctuating gravitational fields. The interaction Hamiltonian, taking the form of Eq.~(\ref{gravitodiam_Ham}), can be expressed as
\beq \label{H_GDE}
H_{AF}+H_{BF}=\frac{1}{2}\hat{\sigma}^{\mathrm{A}}_{ljnf}B_{lj}\left( \textbf{r}_{A} \right)B_{nf}\left( \textbf{r}_{A} \right)+\frac{1}{2}\hat{\sigma}^{\mathrm{B}}_{ljnf}B_{lj}\left( \textbf{r}_{B} \right)B_{nf}\left( \textbf{r}_{B} \right),
\eeq
where $\textbf{r}_{A}$ and $\textbf{r}_{B}$ denote the positions of objects A and B, respectively.

In the transverse-traceless (TT) gauge, the quantized spacetime perturbation $h_{\mu\nu}$ takes the standard form~\cite{Wu2017,Oniga2016}
\beq \label{hij_new}
h_{ij}(\textbf{r},t)
=\sum_{\textbf{k},\lambda} \sqrt{\frac{8\pi G}{(2\pi)^3\omega}} e_{ij}(\textbf{k},\lambda) 
\left[ a_{\textbf{k},\lambda}
e^{i\textbf{k}\cdot\textbf{r}-i\omega t}
+a_{\textbf{k},\lambda}^{\dagger} e^{-i\textbf{k}\cdot\textbf{r}+i\omega t}\right],
\eeq
where $G$ is Newton's gravitational constant, $\textbf{k}$  the wave vector, $\omega=|\textbf{k}|$ the field frequency, $e_{ij}(\textbf{k},\lambda)$ the gravitational polarization tensor, $\lambda$ the polarization states, and $a_{\textbf{k},\lambda}$ and $a_{\textbf{k},\lambda}^{\dagger}$ the annihilation and creation operators of the quantum gravitational field with wave vector $\textbf{k}$ and polarization $\lambda$, respectively.

Within the weak-field approximation, by utilizing the relations between the Riemann curvature tensor and the metric perturbation, and noting that the Riemann tensor coincides with the Weyl tensor in vacuum, the gravitomagnetic field can be formulated as~\cite{hao2024} 
\beq \label{Bij_h}
B_{ij}
=-\frac{1}{2}\epsilon_{ifl}\partial_{f}\dot{h}_{lj},
\eeq
where the dot denotes the first derivative with respect to time. Substituting Eq.~(\ref{hij_new}) into Eq. (\ref{Bij_h}), the quantized gravitomagnetic field takes the form
\begin{align}
 \label{q_Bij_h}
B_{ij}(\textbf{r},t)
\nonumber
=&-\frac{1}{2}\sum_{\lambda}\int d^{3}\textbf{k}\sqrt{\frac{8\pi G\omega^{3}}{(2\pi)^{3}}}\epsilon_{ifl}e^{f}_{3}e_{lj}(\textbf{k},\lambda)  \\
&\times \left[ a_{\textbf{k},\lambda}(t)
e^{i\textbf{k}\cdot\textbf{r}}
+ a_{\textbf{k},\lambda}^{\dagger}(t)
e^{-i\textbf{k}\cdot\textbf{r}}\right],
\end{align}
where $\textbf{e}_{3}=\textbf{k}/|\textbf{k}|$ is the unit vector along the propagation direction of the gravitational field, and $e^{f}_{3}~(f=x,y,z)$ denotes the $f$th coordinate component of $\textbf{e}_{3}$.

\begin{figure}[ht]
\includegraphics[scale=0.4]{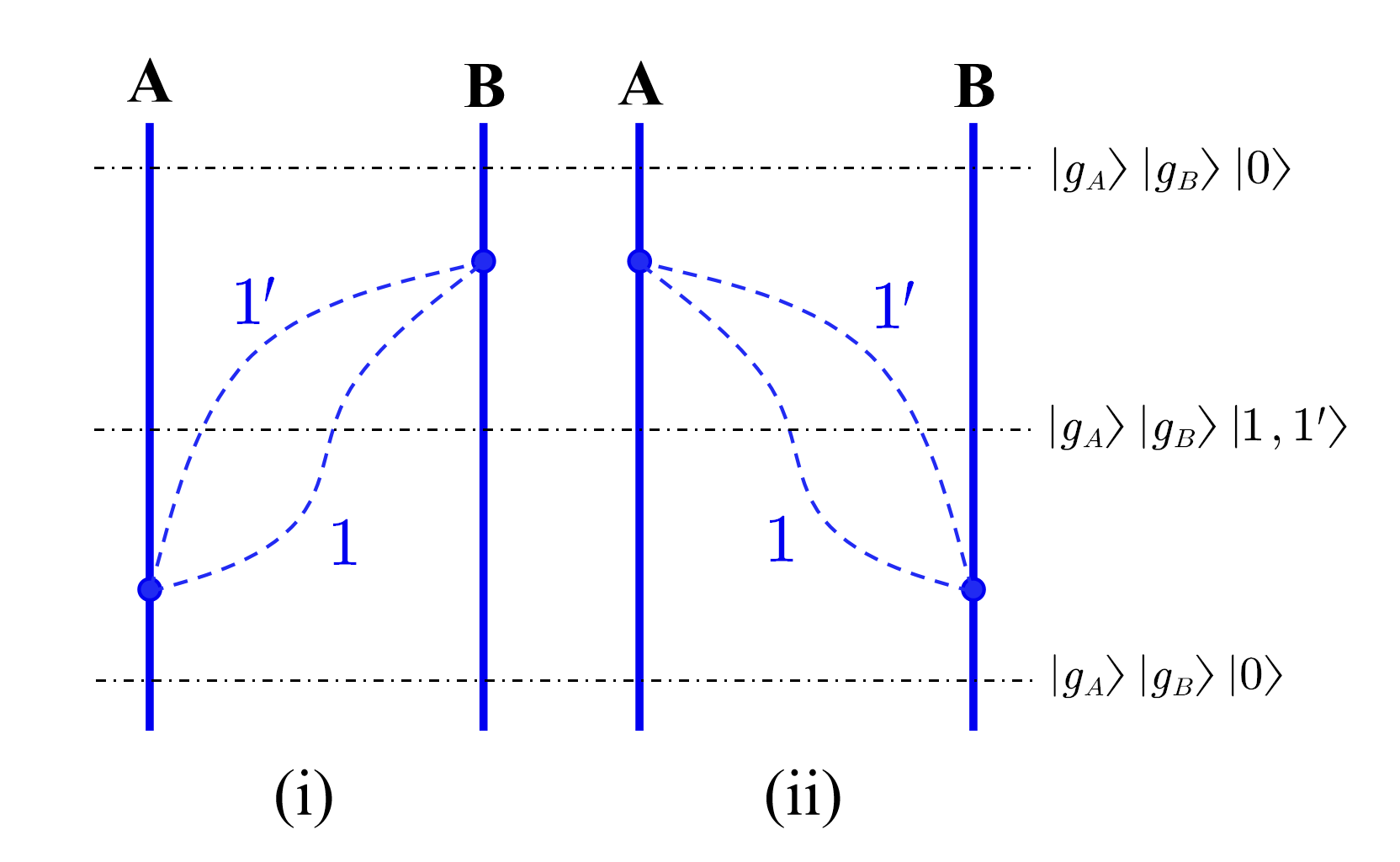}\vspace{0.0cm}
\caption{\label{Feyman_1} 
Time-ordered diagrams contributing to the quantum gravitodiamagnetic interaction.}
\end{figure}

We assume that the object-field system is initially prepared in an uncoupled ground state, where both the objects are in their ground states, $|g_{A}\rangle$ and $|g_{B}\rangle$, and the gravitational field is in the vacuum state $|0\rangle$. Accordingly, the initial state of the total system is $|\phi\rangle=|g_{A}\rangle|g_{B}\rangle|0\rangle$.
In the following, we calculate the quantum interaction between a pair of massive objects using second-order perturbation theory. 
The gravitodiamagnetic interaction, like the gravitoelectric~\cite{Ford2016,Wu2016} and gravitomagnetic~\cite{hao2024} cases, involves two-graviton exchange at fourth order in the gravitational coupling constant. The difference in perturbative order stems from the Hamiltonian structure: While the linear gravitoelectric and gravitomagnetic Hamiltonians require a fourth-order perturbative calculation, the quadratic gravitodiamagnetic Hamiltonian (\ref{H_GDE}) captures the two-graviton contribution already at second order in perturbation theory. The resulting interaction energy is
\beq \label{E_AB}
\Delta E^{\mathrm{GDM}}_{AB}=-\sum_{\mathrm{I}\neq\phi}\frac{\langle \phi|H_{AF}+H_{BF}|\mathrm{I}\rangle\langle \mathrm{I}|H_{AF}+H_{BF}|\phi\rangle}{E_{\mathrm{I}}-E_{\phi}}.
\eeq
Here, $|\mathrm{I}\rangle$ represents the intermediate state in the interaction process. Since the coupling between the object and the gravitomagnetic field is quadratic in the field operator, each object interacts with the field once and then returns to its ground state. This process involves two time-ordered diagrams, as shown in Fig.~{\ref{Feyman_1}}: (i) object A emits two virtual gravitons, which are absorbed by object B, and  (ii) object B emits two virtual gravitons, which are absorbed by object A. Hence, the only relevant intermediate state is $|\mathrm{I}\rangle=|g_A \rangle|g_B \rangle|1,1'\rangle$,
where $|1,1'\rangle$ is the two-graviton state.  Correspondingly, the denominator of Eq.~(\ref{E_AB}) is $E_{\mathrm{I}}-E_{\phi}=\omega +\omega' $. Substituting this intermediate state into Eq.~(\ref{E_AB}), the interaction energy between the two objects becomes
\begin{align} \label{E_GDM_in}
\nonumber
\Delta E_{AB}^{\mathrm{GDM}}\left( \mathbf{r}_A,\mathbf{r}_B \right) =&-\int_0^{+\infty}{d\omega}\int_0^{+\infty}{d\omega'}\left( \frac{1}{\omega +\omega'} \right) \bar{\sigma}_{ljnf}^{A}\bar{\sigma}_{kmsh}^{B}    \\
&\times G_{ljkm}^{\mathrm{MM}}\left( \omega ,\mathbf{r}_A,\mathbf{r}_B \right) G_{nfsh}^{\mathrm{MM}}\left( \omega',\mathbf{r}_A,\mathbf{r}_B \right).
\end{align}
Here, $\bar{\sigma}_{ljnf}^{A\left( B \right)}\equiv \langle g_{A\left( B \right)}|\hat{\sigma}_{ljnf}^{A\left( B \right)}|g_{A\left( B \right)}\rangle$ denotes the ground-state gravitodiamagnetizability of object A(B) (see also the definition in Eq.~(\ref{GDM_Snf})), and $G_{ljkm}^{\mathrm{MM}}\left( \omega ,\mathbf{r}_A,\mathbf{r}_B \right)$ is the two-point correlation function of the gravitomagnetic field in the frequency domain, which takes the form
\beq \label{G_ijab}
G^{\mathrm{MM}}_{ljkm}(\omega, \textbf{r}_{A},\textbf{r}_{B})=\langle 0|B_{lj}(\omega,\textbf{r}_{A})B_{km}(\omega,\textbf{r}_{B})|0\rangle.
\eeq

For simplicity, we assume that both objects A and B are modeled as gravitational hydrogen atoms. Consequently, their ground states are spherically symmetric, implying that their physical properties remain invariant under arbitrary spatial rotations. Consequently, directional dependencies of tensorial quantities should be removed through rotational averaging~\cite{Salam2010,Andrews1977}. Accordingly, the ground-state gravitodiamagnetizability $\bar{\sigma}_{ljnf}^{A\left( B \right)}$, as a fourth-rank tensor, should satisfy the following isotropic condition~\cite{Salam2010,Andrews1977}
\beq \label{R_iso_GDM_ground}
\bar{\sigma}_{ijkl}^{A(B)}=\sum_{\mathbb{P} \in A(B)} \frac{m_{\mathbb{P}} \bar{r}_{\mathbb{P}}^{4}}{135} \left( 4\delta _{ik}\delta _{jl}-\delta _{kj}\delta _{il}-\delta _{kl}\delta _{ji} \right),
\eeq
where $\bar{r}_{\mathbb{P}}$ denotes the expectation value of the distance from mass point $\mathbb{P}$ to the center of mass of object A(B). The detailed derivation of Eq. (\ref{R_iso_GDM_ground}) is shown in ~\ref{appd1}. Substituting Eq.~(\ref{R_iso_GDM_ground}) into Eq.~(\ref{E_GDM_in}), and noting that the gravitomagnetic field $B_{ij}$ is a symmetric traceless tensor, one obtains
\begin{align} \label{E_GDM_in_iso}
\nonumber
\Delta E_{AB}^{\mathrm{GDM}}\left( \mathbf{r}_A,\mathbf{r}_B \right) =&-\frac{1}{25}\int_0^{+\infty}{d\omega}\int_0^{+\infty}{d\omega'}\left( \frac{1}{\omega +\omega'} \right) \sigma_{A}\sigma_{B}    \\
&\times G_{ljkm}^{\mathrm{MM}}\left( \omega ,\mathbf{r}_A,\mathbf{r}_B \right) G_{ljkm}^{\mathrm{MM}}\left( \omega',\mathbf{r}_A,\mathbf{r}_B \right),
\end{align}
where $\sigma _{A\left( B \right)}\equiv \frac{1}{9} \sum_{\mathbb{P} \in A\left( B \right)}{m_{\mathbb{P}}\bar{r}^{4}_{\mathbb{P}}}$.
The frequency-domain two-point correlation function $G^{\mathrm{MM}}_{ljkm}(\omega, \textbf{r}_{A},\textbf{r}_{B})$ in the equation above can be derived from the time-domain one via a Fourier transform. The time-domain two-point correlation function of the gravitomagnetic fields can be obtained by using Eqs. (\ref{q_Bij_h}) and (\ref{G_ijab}) as~\cite{hao2024}
\beq \label{G_ijab_cal}
G^{\mathrm{MM}}_{ljkm}(\textbf{r},{\textbf{r}}',t,{t}')
=\int d^{3}\textbf{k}\frac{G\omega^{3}}{(2\pi)^{2}}\mathcal{G}^{\mathrm{MM}}_{ljkm}(\textbf{k})e^{i\textbf{k}\cdot(\textbf{r}-{\textbf{r}}')-i\omega(t-{t}')},  
\eeq
where $\mathcal{G}^{\mathrm{MM}}_{ljkm}(\textbf{k})$ denotes the gravitomagnetic polarization summation, which takes the form
\begin{align} \label{polar_sum}
\nonumber
\mathcal{G}^{\mathrm{MM}}_{ljkm}(\textbf{k})
\nonumber=&\delta_{lk}\delta_{jm}+\delta_{lm}\delta_{jk}-\delta_{lj}\delta_{km}+\hat{k}_{l}\hat{k}_{j}\hat{k}_{k}\hat{k}_{m}+\hat{k}_{l}\hat{k}_{j}\delta_{km}  \\  
&+\hat{k}_{k}\hat{k}_{m}\delta_{lj}-\hat{k}_{l}\hat{k}_{k}\delta_{jm}-\hat{k}_{l}\hat{k}_{m}\delta_{jk}-\hat{k}_{j}\hat{k}_{k}\delta_{lm}-\hat{k}_{j}\hat{k}_{m}\delta_{lk},
\end{align}
with $\hat{k}_{i}$ being the $i$th component of the unit wave vector $\hat{k}=\textbf{k}/k$. Transforming Eq. (\ref{G_ijab_cal}) into the spherical coordinate system of $\textbf{k}$, one obtains
\begin{align} \label{G_ijab_sph}
\nonumber
G^{\mathrm{MM}}_{ljkm}(r,\Delta t)=&\int_{0}^{+\infty}d\omega\frac{G\omega^{5}}{(2\pi)^{2}}\int_{0}^{\pi}d\theta\sin\theta\int_{0}^{2\pi}d\varphi \mathcal{G}^{\mathrm{MM}}_{ljkm}(\theta,\varphi) \\
&\times e^{i\omega(r\cos\theta-\Delta t)},
\end{align}
where $r=|\textbf{r}-{\textbf{r}}'|$ is the distance between the two objects, and $\Delta t=t-{t}'$. By transforming Eq. (\ref{G_ijab_sph}) into the frequency domain via a Fourier transform, substituting the result into Eq. (\ref{E_GDM_in_iso}), 
and performing the integration over the angular variables $(\theta,\varphi,{\theta}’,{\varphi}’)$, the interaction energy takes the form
\begin{align} \label{E_GDM_simp}
 \nonumber
\Delta E^{\mathrm{GDM}}_{AB}\left( r \right)=&-\frac{4 G^{2}\sigma_A\sigma_B}{25\pi ^2 r^{10}}\int_0^{+\infty}{d\omega}\Bigg\{ \int_0^{+\infty}{d\omega' }\left( \frac{1}{\omega +\omega' } \right)  \\
\nonumber
&\times \left[ A \left( \omega r,\omega'r \right) -i B \left( \omega r,\omega'r \right) \right] e^{i\omega'r }      \\
\nonumber
&+\int_0^{-\infty}{d\omega'}\left( \frac{1}{\omega -\omega'} \right) \\
&\times \left[ A \left( \omega r,\omega'r \right) -i B \left( \omega r,\omega'r \right) \right] e^{i\omega'r} \Bigg\}.
\end{align}
where the polynomials $A\left( x,x' \right)$ and $B\left( x,x' \right)$ are given by
\begin{align} \label{A_1}
\nonumber
A(x,x’)=& xx'\left( 315+8 x^2 x'^2-30 x^2-30 x'^2 \right) \cos x   \\
\nonumber
&+x'\left( -315+135 x^2+30 x'^2-18 x^2 x'^2-3 x^4 \right.  \\
&\left. +2 x^4 x'^2 \right) \sin x,
\end{align}
and
\begin{align} \label{B_1}
\nonumber
B(x,x')=&x\left( -315+135 x'^2+30 x^2-18 x^2 x'^2-3 x'^4 \right. \\
\nonumber
& \left. +2 x^2 x'^4\right) \cos x +\left( 315-135 x^2+3 x^4-135 x'^2 \right. \\
&\left. +63 x^2 x'^2-3 x^4 x'^2+3 x'^4-3 x^2 x'^4+x^4 x'^4 \right) \sin x,
\end{align}
respectively. 
To simplify the integral over $\omega'$, we employ contour integration via the substitution $\omega'=iu$, shifting the integration path to the imaginary axis. This leads to 
\begin{align} \label{E_GDM_simp_3}
\nonumber
\Delta E_{AB}^{\mathrm{GDM}}\left( r \right)=&-\frac{2G^2\sigma _A\sigma _B}{25\pi ^2r^{10}}\int_0^{+\infty}{e^{-ur}du}\int_{-\infty}^{+\infty}{d\omega}  \\
\nonumber
&\times\left( \frac{1}{\omega +iu}+\frac{1}{\omega -iu} \right)    \\
&\times e^{i\omega r}\left[ X\left( \omega r,ur \right) -iY\left( \omega r,ur \right) \right],
\end{align}
where the polynomials $X\left( x,y \right)$ and $Y\left( x,y \right)$ take the form
\begin{align} \label{X_1}
\nonumber
X(x,y)=& -xy\left( 315-8 x^2 y^2-30 x^2+30 y^2 \right)    \\
&+x \left( -315-135 y^2+30 x^2+18 x^2 y^2-3 y^4+2 x^2 y^4 \right),
\end{align}
and
\begin{align} \label{Y_1}
\nonumber
Y(x,y)=&y\left( 315-135 x^2+30 y^2-18 x^2 y^2+3 x^4
+2 x^4 y^2\right)   \\
\nonumber
&+\left( 315-135 x^2+3 x^4+135 y^2-63 x^2 y^2+3 x^4 y^2 \right. \\
&\left. +3 y^4-3 x^2 y^4+x^4 y^4 \right) ,
\end{align}
respectively. Performing the Cauchy principal value integral over the variable $\omega$ in Eq. (\ref{E_GDM_simp_3}), 
we obtain
\beq \label{E_GDM_simp_4}
\Delta E_{AB}^{\mathrm{GDM}}\left( r \right)=-\frac{4G^2\sigma _A\sigma _B}{25\pi r^{10}}\int_0^{+\infty}{du}T\left( ur \right) e^{-2ur},
\eeq
where the polynomial $T\left(y \right)$ reads
\begin{align} \label{Ty}
\nonumber
T\left( y \right) =&315+630y+585y^2+330y^3+129y^4+42y^5  \\
&+14y^6+4y^7+y^8.
\end{align}
Performing the integral over the variable $u$ in Eq.~(\ref{E_GDM_simp_4}), the quantum gravitodiamagnetic interaction potential energy is found to be
\beq \label{E_GDM_simp_5}
\Delta E_{AB}^{\mathrm{GDM}}\left( r \right)=-\frac{3987\hbar c G^2}{25\pi r^{11}}\sigma_A \sigma_B.
\eeq
Note that the result above is expressed in International System of Units (SI units), 
in which an additional factor of ${1}/{c^{2}}$ must be included in $\sigma_{A(B)}$. 
This $r^{-11}$ scaling contrasts with the power-law behavior reported in previous studies of quantum gravitational quadrupole interactions between two polarizable objects induced by vacuum fluctuations of the gravitational field~\cite{hao2024,Ford2016,Wu2016,hao2025,Holstein2017}. In those works, the interaction scales as $r^{-10}$~\cite{hao2024,Ford2016,Wu2016,Holstein2017} or $r^{-8}$~\cite{hao2025} in the near regime, and as $r^{-11}$ in the far regime. The origin of this difference is explained below.

For quantum gravitational quadrupole interactions, such as the gravitomagnetic interaction, the general form of the interaction potential energy is expressed as~\cite{hao2024}
\beq \label{E_AB_GM}
\Delta E^{\mathrm{GM}}_{AB}(r)\propto -\frac{1}{r^{10}}\int_{0}^{+\infty}du\chi_{A}(iu)\chi_{B}(iu)T(ur)e^{-2ur},
\eeq
where
$T(x)$ is shown in Eq.~\eqref{Ty}, $\chi_{A(B)}(iu)$
is the ground-state gravitomagnetizability of object A(B), and $iu=\omega$ is the frequency.
Notably, this expression contains an integral over the variable $u$, with both the gravitomagnetizability $\chi_{A(B)}\left( iu \right)$ and the polynomial $T(ur)$ being explicit functions of $u$. This configuration leads to distinct asymptotic behaviors depending on the interobject separation: When the separation $r$ is much smaller than the transition wavelength of the objects $\omega_{A(B)}^{-1}$, the $ur$-dependent terms in polynomial $T(ur)$ become negligible, leaving only the constant term $T(0)=315$. Consequently, the integral yields no $r$-dependent terms, resulting in a potential scaling as $r^{-10}$. On the other hand, when the separation between the two objects is much larger than the transition wavelength of the objects, the dominant contribution to the integral arises from small values of $u$ due to the exponential suppression factor $e^{-2ur}$. Then the gravitomagnetizability $\chi_{A(B)}(iu)$ can be approximated by its static value $\chi_{A(B)}(0)$. In this case, the polarizabilities can be taken outside the integral, while all $ur$-dependent terms $T(ur)$ should be retained. This ultimately introduces an additional $r^{-1}$ term from the integral, resulting in a long-range potential scaling as  $r^{-11}$. 
Returning to our current work, the ground-state gravitodiamagnetizability $\bar{\sigma}_{ijkl}^{A\left( B \right)}$ of the  objects, as shown in Eq.~(\ref{E_GDM_in}), does not explicitly contain the integration variable $u$ (where $u=\omega'/i$) and can be taken outside the integral as a constant. This means that, irrespective of whether the separation lies in the near or far regime, the gravitodiamagnetizability is a constant and remains equal to its static value. 

Eq.~(\ref{E_GDM_simp_5}) is obtained within the leading Fermi-coordinate approximation. Let $a$ denote the common characteristic size of the two objects, with internal Fermi coordinates satisfying $|\mathbf{x}|\lesssim a$. The higher-order Fermi-coordinate expansions of both $h_{0i}$ and $h_{00}$ introduce curvature-gradient terms containing additional internal coordinates and spatial derivatives. In the two-object interaction, these derivatives act on the gravitomagnetic two-point correlation function $G_{ijkl}^{\mathrm{MM}}(\mathbf r,\Delta t)$, where $\mathbf r=\mathbf r_A-\mathbf r_B$. The scaling behavior of this correlation function implies that a spatial derivative carries one additional inverse power of $r$ in a scaling analysis. Since the internal coordinates are bounded by $a$, the resulting finite-size corrections are generically of relative order $a/r$. Thus, Eq.~(\ref{E_GDM_simp_5}) is valid within the regime $a\ll r$. A similar finite-size restriction was imposed in Ref.~\cite{Ford2016}, although a different approach was employed.

Finally, for the gravitationally bound system considered here, 
we now provide explicit order-of-magnitude estimates of the gravitodiamagnetic interaction and compare it with the leading quantum gravitational quadrupole interaction, the gravitoelectric-gravitoelectric interaction.
For two objects, each composed of two equal-mass particles ($m_{1}=m_{2}=m_0$) in circular orbits of radius $R_0$ bound purely by gravity, the gravitodiamagnetizabilities are $\sigma_A=\sigma_B=\tfrac{2}{9}m_0R_0^4$, and the corresponding quantum gravitodiamagnetic potential scales as
\beq \label{gdm_int_h}
\Delta E_{AB}^{\mathrm{GDM}}\left( r \right) \sim-\frac{\hbar c}{r}\left( \frac{R_0}{r} \right) ^8\left( \frac{R_{\mathrm{S}}}{r} \right) ^2,
\eeq
where $R_{\mathrm{S}}=2Gm_{0}/c^{2}$ is the Schwarzschild radius of the particles.
For comparison, 
the quantum gravitoelectric-gravitoelectric interaction in the far and near regimes, assuming  the same gravitoelectric polarizability $\alpha _A\left( 0 \right) =\alpha _B\left( 0 \right) =\frac{m_{0}R_{0}^{2}}{\omega _{0}^{2}}$ (where $\omega _0=1/2\sqrt{Gm_{0}/R_{0}^{3}}$ denotes the orbital angular frequency) for the two objects,  scales as $\Delta E_{AB}^{\mathrm{GE},\mathrm{far}}\left( r \right) \sim -\left( \hbar c /r \right)\left( R_0/r \right) ^{10}$ and $\Delta E_{AB}^{\mathrm{GE},\mathrm{near}}\left( r \right) \sim -\hbar \omega _0\left( R_0/r \right) ^{10}$~\cite{Ford2016}, respectively.  
For the Newtonian equal-mass binary considered here, the orbital velocity $v=\omega_0R_0$ satisfies $v^2/c^2=(\omega_0R_0)^2/c^2=R_S/(8R_0)\ll1$. Thus, the weak-field and nonrelativistic approximations require $R_S/R_0\ll1$. Moreover, the characteristic size $a$ of each binary is of the same order as $R_0$. The finite-size condition $a\ll r$ therefore implies $R_0/r\ll1$, so that the relevant length scales satisfy $R_S\ll R_0\sim a\ll r$.
Taking the ratios of the above estimates gives $\left|
\Delta E_{AB}^{\mathrm{GDM}}(r)/
     \Delta E_{AB}^{\mathrm{GE,far}}(r)
\right|
\sim
\left(R_S/R_0\right)^2
\ll1$
and
$\left|
\Delta E_{AB}^{\mathrm{GDM}}(r)/
     \Delta E_{AB}^{\mathrm{GE,near}}(r)
\right|
\sim
(R_S/r)
\left(R_S/R_0\right)^{1/2}
=
(R_0/r)
\left(R_S/R_0\right)^{3/2}
\ll1$, respectively. Thus, within the regime of validity of the present model, the gravitodiamagnetic interaction remains parametrically smaller than the leading gravitoelectric--gravitoelectric interaction in both the far and near regimes.

\section{Summary }
\label{sec_disc}

In this work, within the framework of Weyl gravitoelectromagnetism, we derive the fundamental form of the gravitodiamagnetic coupling Hamiltonian starting from the Lagrangian of a particle in a gravitational field. Our analysis reveals that, unlike linear couplings such as gravitoelectric and gravitomagnetic couplings, the gravitodiamagnetic coupling exhibits a quadratic dependence on the gravitomagnetic field. 
Based on this Hamiltonian, we further demonstrate that, for a spherically symmetric gravitational hydrogen-like system in its ground state, the induced quadrupole moment  has the  opposite sign to the applied gravitomagnetic field, which is the defining signature  of gravitodiamagnetism.
Using  linearized quantum gravity, we then investigate the quantum gravitational interaction arising from the  coupling of two massive objects to fluctuating gravitomagnetic  vacuum fields  through their gravitodiamagnetic response. With the help of the leading-order perturbation theory, we derive an explicit expression for the gravitodiamagnetic interaction potential, valid for $r\gg a$, where $r$ is the separation and $a$ the characteristic size of the objects. We find that the resulting quantum gravitodiamagnetic interaction potential is attractive and scales as $r^{-11}$ throughout its regime of validity. 
For the Newtonian binary model used in our order-of-magnitude estimates, this contribution remains parametrically smaller than the leading gravitoelectric-gravitoelectric interaction in both the near and far regimes within the regime of validity of the model.

\section*{Acknowledgements}
This work was supported in part by the NSFC under Grants No. 12635002 and 12575051, the innovative research group of Hunan Province under Grant No. 2024JJ1006, and Postgraduate Scientific Research Innovation Project of Hunan Province under Grant No. CX20250748.

\appendix
\section{Gravitodiamagnetizability and the induced gravitomagnetic quadrupole moment of an isotropic ground-state object}\label{appd1}

Substituting the explicit expression of the gravitodiamagnetizability operator $\hat{\sigma}_{ljnf}^{A\left( B \right)}$ shown in Eq. (\ref{gravitodiam_Ham}) into $\bar{\sigma}_{ljnf}^{A\left( B \right)}\equiv \langle g_{A\left( B \right)}|\hat{\sigma}_{ljnf}^{A\left( B \right)}|g_{A\left( B \right)}\rangle$,  we obtain the ground-state gravitodiamagnetizability as
\beq \label{Ap_1}
\bar{\sigma}_{ljnf}^{A\left( B \right)}=\frac{1}{9}\sum_{\mathbb{P} \in A\left( B \right)}{m_{\mathbb{P}}\left[ \bar{r}_{\mathbb{P}}^{2} \left< x_{j}^{\mathbb{P}}x_{f}^{\mathbb{P}} \right> \delta _{ln}-\left< x_{n}^{\mathbb{P}}x_{j}^{\mathbb{P}}x_{l}^{\mathbb{P}}x_{f}^{\mathbb{P}} \right> \right]},
\eeq
where $\bar{r}_{\mathbb{P}}$ denotes the expectation value of the distance from mass point $\mathbb{P}$ to the center of mass of object A(B). Note that the terms  
$\left< x_{j}^{\mathbb{P}}x_{f}^{\mathbb{P}} \right>$ and $\left< x_{n}^{\mathbb{P}}x_{j}^{\mathbb{P}}x_{l}^{\mathbb{P}}x_{f}^{\mathbb{P}} \right>$ in Eq. (\ref{Ap_1}) represent the Cartesian components of a second-rank tensor and a fourth-rank tensor, respectively, in the laboratory frame. Their counterparts in the frame where the mass points are fixed are denoted as $\left< x_{a}^{\mathbb{P}}x_{b}^{\mathbb{P}} \right>$ and $\left< x_{a}^{\mathbb{P}}x_{b}^{\mathbb{P}}x_{c}^{\mathbb{P}}x_{d}^{\mathbb{P}} \right>$, respectively. According to the rotational averaging method for isotropic systems, the second- and fourth-rank tensors in the two reference frames are related by the following expressions $\left< x_{j}^{\mathbb{P}}x_{f}^{\mathbb{P}} \right> =\frac{1}{3}\delta _{jf} \bar{r}_{\mathbb{P}}^{2}$ and $\left< x_{n}^{\mathbb{P}}x_{j}^{\mathbb{P}}x_{l}^{\mathbb{P}}x_{f}^{\mathbb{P}} \right> 
=\frac{\bar{r}_{\mathbb{P}}^{4}}{15}\left( \delta _{nj}\delta _{lf}+\delta _{nl}\delta _{jf}+\delta _{nf}\delta _{jl} \right)$~\cite{Salam2010,Andrews1977}.
Substituting the two expressions into Eq. (\ref{Ap_1}), one obtains
\beq \label{AB_gra_dia_iso_2}
\bar{\sigma}_{ijkl}^{A(B)}=\sum_{\mathbb{P} \in A(B)} \frac{m_{\mathbb{P}} \bar{r}_{\mathbb{P}}^{4}}{135} \left( 4\delta _{ik}\delta _{jl}-\delta _{kj}\delta _{il}-\delta _{kl}\delta _{ji} \right),
\eeq
which is Eq. \eqref{R_iso_GDM_ground} in the paper.
To highlight the characteristic behavior of gravitodiamagnetism, we calculate the induced gravitomagnetic quadrupole moment of a ground-state gravitational hydrogen atom with spherical symmetry based on Eq. (\ref{GDM_Snf}). Substituting Eq.~(\ref{AB_gra_dia_iso_2}) into Eq.~\eqref{GDM_Snf}, and noting that the gravitomagnetic field $B_{ij}$ is a symmetric traceless tensor, the induced gravitomagnetic quadrupole moment reads
\beq \label{GDM_Snf_App}
S_{nf}=-\frac{1}{30}\sum_{\mathbb{P} \in A(B)}m_{\mathbb{P}} \bar{r}_{\mathbb{P}}^{4} B_{nf},
\eeq
where $\sum_{\mathbb{P} \in A(B)} m_{\mathbb{P}} \bar{r}^{4}_{\mathbb{P}}$ is positive. This clearly shows that the gravitomagnetic field always induces a quadrupole moment with the opposite sign, which is a defining characteristic of gravitodiamagnetism.

\end{document}